\documentclass[aps, pra, reprint, floatfix, superscriptaddress]{revtex4-1} 

\usepackage{graphicx}
\usepackage{epstopdf}
\usepackage{bm, amsmath, amssymb}
\usepackage{float}
\usepackage{placeins}
\usepackage{xcolor}
\usepackage{siunitx}
\usepackage[T1]{fontenc}
\usepackage{array,booktabs}
\usepackage{comment}
\usepackage{multirow}
\usepackage{enumitem} 

\newcommand{\figwidth}{3.375in} 
\newcommand{\figwidthDouble}{7in}

\begin{document}

\title{Josephson Array Mode Parametric Amplifier}

\author{V. V. Sivak}
\email{vladimir.sivak@yale.edu}
\affiliation{Department of Applied Physics, Yale University, New Haven, CT 06520, USA}

\author{S. Shankar}
\thanks{Current address: Department of Electrical and Computer Engineering, University of Texas at Austin, Austin, TX 78712, USA}
\affiliation{Department of Applied Physics, Yale University, New Haven, CT 06520, USA}

\author{G. Liu}
\affiliation{Department of Applied Physics, Yale University, New Haven, CT 06520, USA}

\author{J. Aumentado}
\affiliation{National Institute of Standards and Technology, Boulder, CO 80305, USA}

\author{M. H. Devoret}
\affiliation{Department of Applied Physics, Yale University, New Haven, CT 06520, USA}

\begin{abstract}
We introduce a novel near-quantum-limited amplifier with a large tunable bandwidth and high dynamic range -- the Josephson Array Mode Parametric Amplifier (JAMPA). The signal and idler modes involved in the amplification process are realized by the array modes of a chain of 1000 flux tunable, Josephson-junction-based, nonlinear elements. The frequency spacing between array modes is comparable to the flux tunability of the modes, ensuring that any desired frequency can be occupied by a resonant mode, which can further be pumped to produce high gain. We experimentally demonstrate that the device can be operated as a nearly quantum-limited parametric amplifier with $20\rm\, dB$ of gain at almost any frequency within $(4-12)\,\rm GHz$ band. On average, it has a $3\rm\, dB$ bandwidth of $11\rm\, MHz$ and input $1\rm\, dB$ compression power of $-108\rm\, dBm$, which can go as high as $-93\rm\, dBm$. We envision the application of such a device to the time- and frequency-multiplexed readout of multiple qubits, as well as to the generation of continuous-variable cluster states.
\end{abstract}

\maketitle

\section{Introduction\label{sec:Introduction}}

Quantum-limited parametric amplifiers have become the necessary tools for quantum measurements involving readout at microwave frequency. The improved signal-to-noise ratio brought about by these amplifiers is crucial in applications such as real-time feedback in quantum error correction \cite{Ofek2016,Campagne-Ibarcq2019} but also simply in quantum benchmarking experiments such as qubit lifetime characterization in superconducting circuits \cite{Houck2008,Burnett2019}. These devices are also used as sources of squeezing for enhancement of detection sensitivity \cite{Eddins2018,Eddins2018a}, and sources of entanglement in continuous variable quantum computing \cite{Flurin2012,SandboChang2018}.

The most common implementations of nearly quantum-limited parametric amplifiers are based on nonlinear resonators containing Josephson junctions, such as Josephson parametric amplifiers (JPA) \cite{Castellanos-Beltran2008, Yamamoto2008} and converters (JPC) \cite{Bergeal2010a, Abdo2013}. These resonant amplifiers have limited bandwidth and dynamic range, and work in reflection, therefore requiring commercial circulators for signal routing. In contrast, the traveling wave parametric amplifiers (TWPA) based on Josephson junctions \cite{OBrien2014,Macklin2015,Bell2015,White2015,Zorin2016,Planat2019} or high-kinetic-inductance materials \cite{HoEom2012,Vissers2015,Chaudhuri2017,Ranzani2018,Zobrist2019} are broad-band, have higher dynamic range, and are, in principle, directional. In recent years, Josephson TWPAs have become practical and widely adopted in the field \cite{Heinsoo2018,Eddins2018,Bultink2018,Krinner2019,Peronnin2019}, despite the fabrication challenges involved in attaining impedance- and resonant phase-matching.

However, TWPAs have failed to achieve one of their original promises, namely directionality. For high coherence superconducting qubits, an extremely clean electromagnetic environment is crucial. This would be compromised if the TWPAs were used without an isolator at the input port, since the reflected strong pump and amplified signal and quantum noise would travel back to the qubits and cause unwanted backaction. Such reflections happen due to small impedance mismatches which are inevitably present within the large bandwidth of the TWPA. In practice, to operate the TWPA, isolators are also essential for impedance matching the input port. Operating a TWPA in such a setting is therefore equivalent to operating a broad-band high-dynamic-range {\it reflection amplifier}. 

This observation naturally leads to the question: by dropping the directionality constraint, is it possible to construct a reflection amplifier which would avoid the two most strict and challenging requirements for the TWPA (i) impedance matching and (ii) phase matching, and at the same time be significantly better than commonplace resonant reflection amplifiers in terms of the bandwidth and dynamic range? 

In this paper, we achieve a more modest goal as a step toward answering this question. We construct an amplifier with the following characteristics: (i) its flux-tunable bandwidth is larger than the dynamic bandwidth of the Josephson TWPA, (ii) its dynamic range is comparable to that of the Josephson TWPA with $P^{\rm average}_{1\rm dB}=-108\rm\, dBm$ and $P^{\rm best}_{1\rm dB}=-93\rm\, dBm$, and (iii) its added noise performance matches that of state-of-the-art near-quantum-limited resonant amplifiers. The basic idea behind the device is to harness the collection of modes of a nonlinear Fabry-Perot resonator, realized as a long array of flux-tunable nonlinear Josephson mixing elements, hence the name Josephson Array Mode Parametric Amplifier (JAMPA). If the frequency spacing between array modes is smaller than the flux tunability of the modes, then any desired frequency can be occupied by a resonant mode at a certain flux, which can further be pumped to produce gain. In practice, the tunable bandwidth of such a device is essentially limited by the $(4-12\rm)\, GHz$ bandwidth of our cryogenic measurement chain.

To get access to three-wave mixing and potential {\it in-situ} Kerr suppression \cite{Sivak2019}, we chose to use an array of SNAILs (superconducting nonlinear asymmetric inductive element \cite{Frattini2018}), although it is possible to make a JAMPA with other types of flux-tunable mixing elements, such as dc- or rf-SQUIDs. We note that in parallel with this work a similar idea was developed in \cite{Winkel2019}, where pairs of modes (dimers) of two coupled arrays of SQUIDs are utilized for amplification.

The remaining of this paper is organized as follows: in Section~\ref{sec:Effect of the array size} we study the effect of the array size on various amplifier properties, demonstrating how increasing the size naturally leads to the idea of the JAMPA. In Section~\ref{sec:Probing the array modes} we present experimental comparison of several devices with array size of 20, 200 and 1000. In Section~\ref{sec:Amplifier characterization} we characterize the JAMPA performance. We conclude and discuss possible applications and future directions in Section~\ref{sec:Conclusion}.

\section{Effect of the array size on amplifier mode structure and nonlinearity\label{sec:Effect of the array size}}

To approach the question posed in the Introduction, we start by exploring what phenomena limit the dynamic range in Josephson parametric amplifiers and how these effects can be mitigated. The two dominant physical mechanisms that are understood to cause the saturation of parametric amplifiers are (i) signal-induced Stark shifts that detune the amplifier from its operating point and (ii) depletion of pump photons due to signal amplification. The former was shown to be the dominant factor in saturation of resonant amplifiers \cite{Liu2017,Frattini2018,Planat2018,Sivak2019}, and some evidence indicates that it might also cause the saturation of TWPAs \cite{Planat2019,OBrien2019}. The spurious Stark shift originates from the quartic Kerr nonlinearity of the resonator hosting the amplification process, and therefore the path towards improved dynamic range requires the suppression of this nonlinearity while still preserving the mixing capability. One promising direction for suppression of Kerr, suggested in Ref. \cite{Eichler2013}, is to replace the single mixing element, such as a Josephson junction, with an array, to distribute the power handling load among many junctions. In this section, we revisit this idea of arraying.

\begin{figure}[t]
 \includegraphics[width = \figwidth]{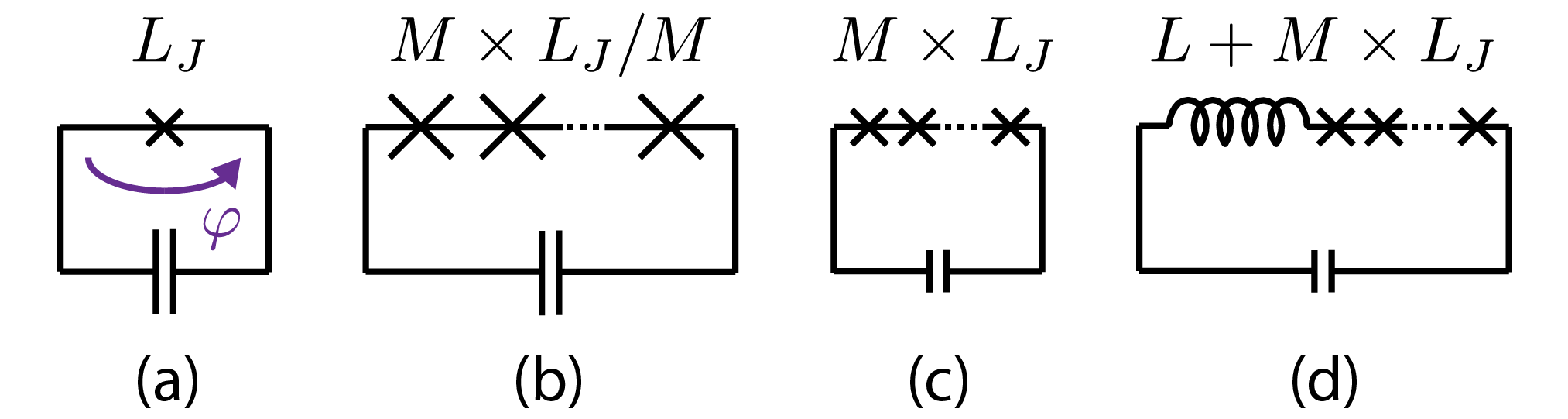}
 \caption{\label{fig1} (a) A simple circuit model of the JPA resonator: junction with inductance $L_J$ shunted with a capacitance $C$. (b) Arraying of $M$ junctions with $M$ times smaller inductance. (c) Arraying of $M$ identical junctions requires smaller capacitance to keep the resonance frequency constant. (d) Array in series with geometric inductance.}
\end{figure}

Consider first a simple model of a JPA resonator, shown in Fig.~\ref{fig1}(a), consisting of a single Josephson junction with the Josephson energy $E_J$ (inductance $L_J$), capacitively shunted with capacitance $C$. Let $\varphi=\varphi_{\rm ZPF}(a+a^\dagger)$ denote the superconducting phase across the junction, where $\varphi_{\rm ZPF}=\sqrt{Z_a/2R_q}$ is the amplitude of zero-point fluctuations (ZPF) of the phase, defined by the ratio of the mode impedance $Z_a$ to the superconducting resistance quantum $R_q=\hbar/(2e)^2$. The Hamiltonian of such a circuit is ${\cal H} = (2eN)^2/2C-E_J\cos\varphi$, where $N$ is the number of Cooper pairs that have tunneled across the junction. We consider the limit of small $\varphi_{\rm ZPF}$ where the cosine expansion is valid, and truncate it at the fourth order obtaining a weakly anharmonic oscillator with resonant frequency $\omega_a$.

The quartic contribution $E_J\varphi^4/4!$ to the Hamiltonian, also referred to as Kerr nonlinearity, can be diluted by replacing a single junction with an array of $M$ junctions each having $M$ times higher Josephson energy ($M$ times smaller inductance), as shown in Fig.~\ref{fig1}(b). Then in the lowest energy configuration the phase $\varphi$ splits equally among the junctions and the quartic contribution to the Hamiltonian becomes
\begin{align}
\frac{M E_J}{4!}\sum_{i=1}^M\left(\frac{\varphi}{M}\right)^4=\frac{E_J}{4!}\frac{\varphi_{\rm ZPF}^4}{M^2}(a^\dagger+a)^4\propto\frac{1}{M^2}. \label{propto1/M^2}
\end{align}
Importantly, in such approach to arraying, the resonant frequency and the impedance of the JPA remain independent of $M$.

\begin{figure*}[t]
 \includegraphics[width = \figwidthDouble]{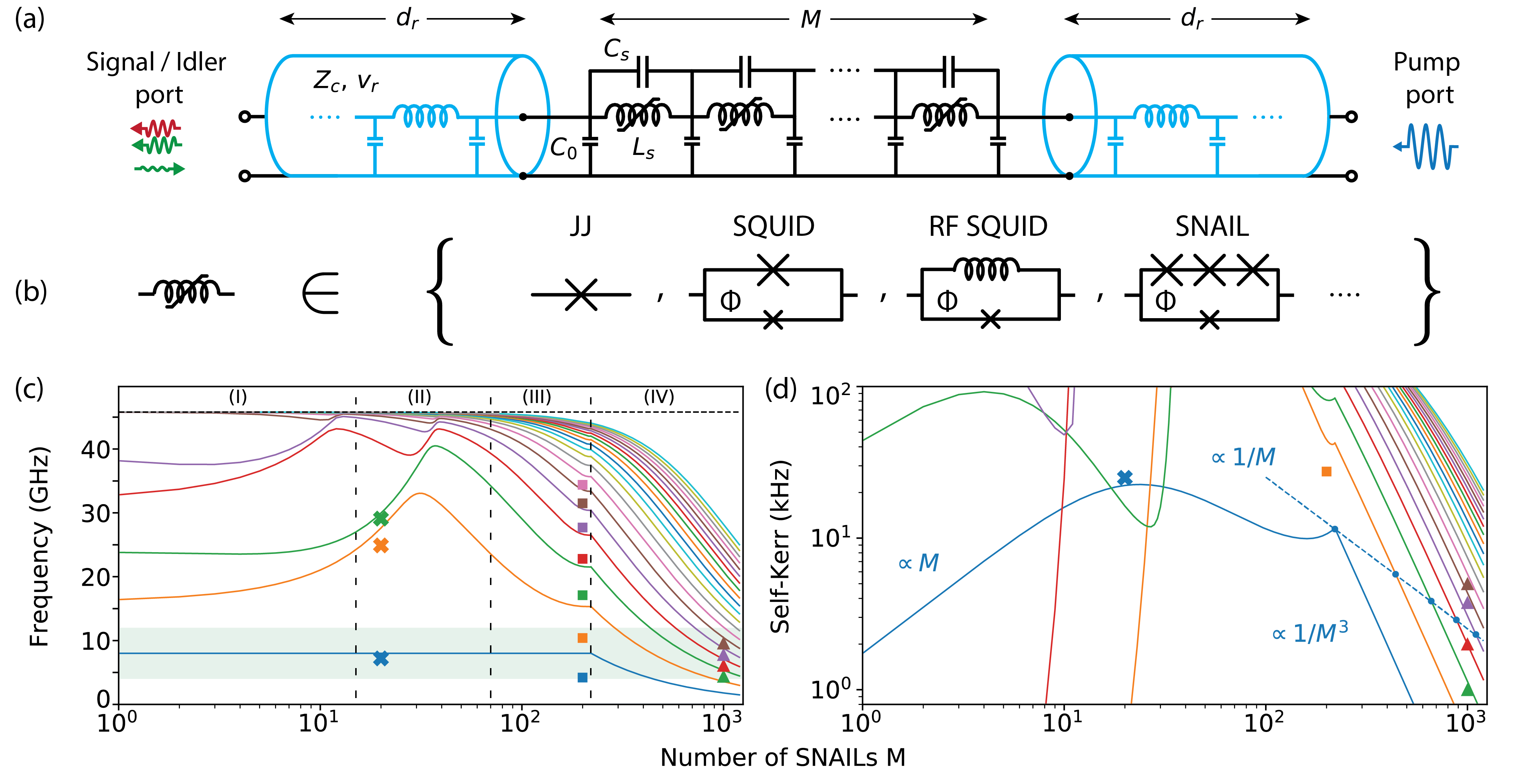}
 \caption{\label{fig2} (a) Distributed circuit model of the type of parametric amplifier considered in this work. An array of $M$ unit cells, comprising of the nonlinear inductance $L_S$ and capacitance $C_0$ and $C_S$, embedded into a transmission line resonator of characteristic impedance $Z_c$ and phase velocity $v_r$. (b) Superconducting circuit elements that can provide nonlinear inductance $L_S$. (c) Resonant frequencies and (d) self-Kerr nonlinearities of the modes of this structure with SNAIL at $\Phi=0$ as nonlinear inductor. For each $M$ the resonator length $2d_r$ is adjusted to keep the fundamental mode of the structure at a fixed frequency. Past a certain critical $M$ the resonator leads shrink to size zero (region IV). The solid curves are calculated for the parameters of the system which are typical for our fabrication method, see Appendix~\ref{App:Eigenmode decomposition}; their color encodes the mode number. Data points correspond to devices A (cross), B (square) and C (triangle) and are shown for reference, not expected to precisely line up with the theory curves.  The $(4-12)\rm\, GHz$ strip denotes the accessible frequency window in which modes can be directly identified using the reflection measurement. Two-tone spectroscopy is used to locate the resonances outside this window. Self-Kerr nonlinearities in (d) are measured using IMD spectroscopy (see text). Vertical dashed lines in (c) denote approximate boundaries where qualitative changes occur in the mode structure of the amplifier.}
\end{figure*}

In practice, however, this approach is not feasible due to the difficulty in fabricating junctions with increasingly larger Josephson energy $ME_J$ as the array size is increased. A more realistic, practical approach is to array fixed-$E_J$ junctions with the largest critical current that can be reliably reproduced in a fabrication process, as shown in Fig.~\ref{fig1}(c). In this case the mode inductance grows $\propto M$ (if the stray geometric inductance can be neglected), and therefore the capacitance has to be adjusted $\propto 1/M$ in order to keep the working frequency the same. This, however, leads to $\propto M$ increase in the mode impedance and corresponding $\propto\sqrt{M}$ increase in $\varphi_{\rm ZPF}$. The last factor significantly affects the ability to dilute the Kerr nonlinearity:
\begin{align}
\frac{E_J}{4!}\sum_{i=1}^M\left(\frac{\varphi}{M}\right)^4=\frac{E_J}{4!}\frac{\varphi_{\rm ZPF}^4}{M^3}(a^\dagger+a)^4\propto\frac{1}{M}.\label{propto1/M}
\end{align}

Furthermore, in practice there is often a linear inductance $L$ in series with the array, as shown in Fig.~\ref{fig1}(d). This can be, for example, the stray geometric inductance of the leads or the intentionally fabricated inductance of the embedding circuit. For junctions with large critical current such that $p_J\equiv L_J/L\ll1$ and small array size such that $Mp_J\ll1$, the phase drop across each junction $\varphi\times p_J/(1+Mp_J)\approx p_J\varphi$ is independent of $M$. In such arrays the Kerr nonlinearity actually increases $\propto M$, as shown below:
\begin{align}
\frac{E_J}{4!}\sum_{i=1}^M(p_J\varphi)^4=\frac{E_J}{4!}M(p_J\varphi_{\rm ZPF})^4(a^\dagger+a)^4\propto M.\label{propto M}
\end{align}

These simple scaling arguments demonstrate that larger array size $M$ is not strictly equivalent to smaller Kerr nonlinearity. To be more quantitative in support of this claim, we consider an analytical model that reflects a practically feasible approach to arraying of nonlinear elements and avoids multiple approximations involved in Eqs.~\eqref{propto1/M^2}-\eqref{propto M}.  In this model, a unit cell of the array is comprised of a nonlinear inductance $L_S$, capacitance to ground $C_0$, and shunting capacitance $C_S$, as shown in Fig.~\ref{fig2}(a). The inductance $L_S$ can come from any superconducting circuit element, such as a Josephson junction, a SQUID, a SNAIL etc, as shown in Fig.~\ref{fig2}(b). The array is embedded into a transmission line resonator, isolated from the environment for analysis simplicity. The two arms of the transmission line resonator are of length $d_r$ each, and have the characteristic impedance $Z_c$ and phase velocity $v_r$.

As a distributed element, this system can support multiple standing electromagnetic modes. When increasing $M$, the length $d_r$ is adjusted to keep the resonant frequency of the fundamental mode fixed. These conditions would reflect the real constraints encountered in amplifier design. For this model, in Appendix~\ref{App:Eigenmode decomposition} we perform the eigenmode decomposition and find the resonant frequency and self-Kerr nonlinearity of each mode as a function of the array size $M$. The result of such calculation is plotted in Fig.~\ref{fig2}{(c) and (d) for realistic system parameters and with the choice of nonlinear inductance as a SNAIL at $\Phi=0$. For numerical analysis, we use the fundamental mode frequency of $8\rm\, GHz$, SNAIL linear inductance $L_S=110\rm\, pH$, distributed capacitance $C_0=0.71\rm\, fF$ and  $C_S=0.11\rm\, pF$, and the resonator characteristic impedance $Z_c=46\rm\, \Omega$ and phase velocity $v_r=1.2\times 10^8\rm\, m/s$.

As seen from Fig.~\ref{fig2}(c), the resonant frequency of the fundamental mode stays constant for $M<220$ by construction.  At small $M$ in region I the higher modes of the structure approximately follow the dimensional quantization law for the $\lambda/2$-type transmission line resonator, where resonances are arranged with the spacing approximately equal to the frequency of the fundamental mode. The higher resonances bunch up tightly below the ``plasma frequency'' of the unit cell $\omega_p=1/\sqrt{L_SC_S}$ \cite{Masluk2012,Manucharyan2012}. The self-Kerr nonlinearity, as seen from Fig.~\ref{fig2}(d), grows like $M$ in agreement with the qualitative discussion leading to Eq.~\eqref{propto M}. In the region II of intermediate $M$, where the inductive energy participation ratio (EPR) \cite{Minev2019} of the array saturates to $1$, the array acts as a superinductance \cite{Masluk2012, Manucharyan2012} and the transmission line leads only contribute to the capacitance. In this case the resonator is well-approximated as a single lumped $LC$-type resonance with all higher modes being pushed up in frequency. Further increase of $M$ in region III results in the growing contribution of the array distributed capacitance $C_0$, and the array modes descend in frequency. The scaling of the Kerr nonlinearity with $M$ in regions II and III changes to approximately $1/M^\gamma$ with $\gamma\approx0.6$ for our choice of parameters, in contrast with the simplistic prediction of Eq.~\eqref{propto1/M} due to the increased role of the array distributed capacitance. Eventually, after the critical $M=220$ is reached, the resonator leads shrink to zero, and we lose the knob that ensures that the frequency of the fundamental mode remains constant. For $M>220$, corresponding to region IV, increasing the array size results in $1/M$ scaling of the mode spacing. The self-Kerr nonlinearity of the fundamental mode is very efficiently diluted as $1/M^3$ in the region IV, but its resonant frequency is pulled down, and therefore the fundamental mode can not be used if the desired operating frequency is fixed. In this case one would have to use higher array modes that occupy the desired operating frequency. Taking this into account, the effective scaling of self-Kerr nonlinearity goes only as $1/M$, as depicted in Fig.~\ref{fig2}(c) with a dashed line.

\begin{figure*}
 \includegraphics[width = \figwidthDouble]{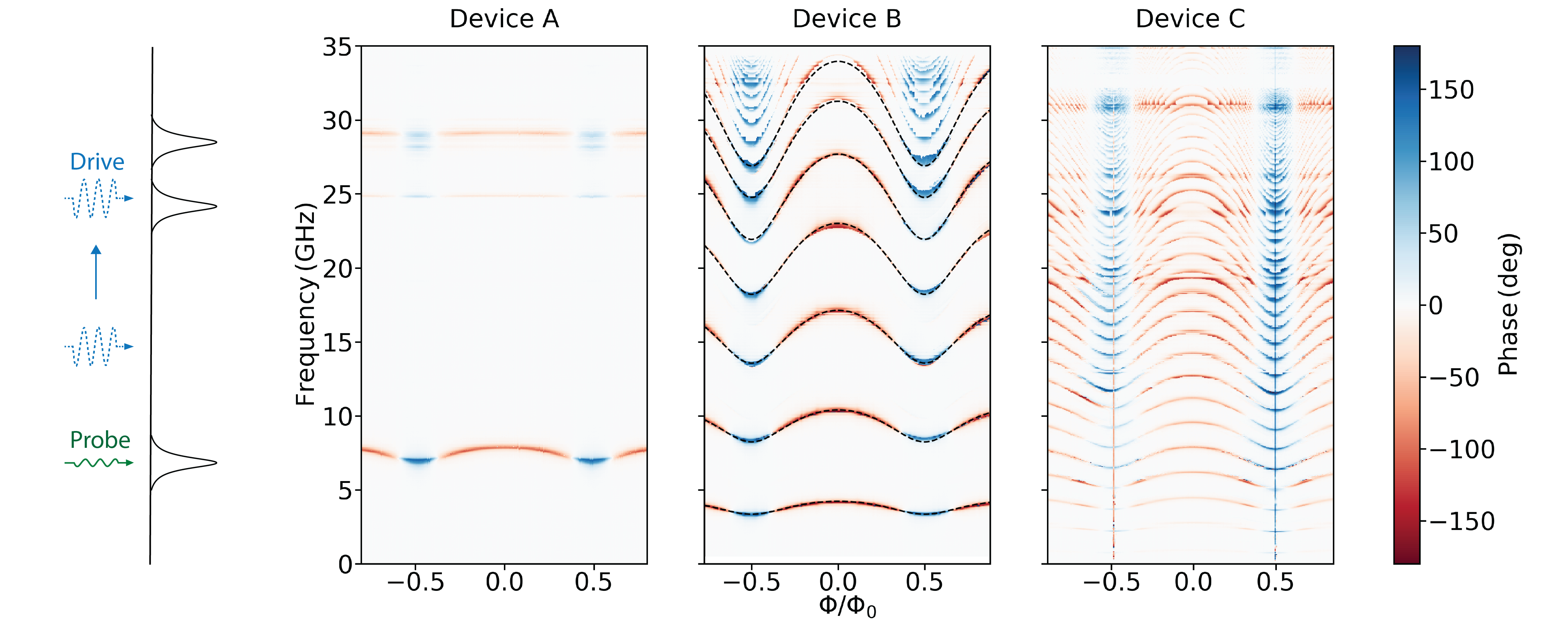}
 \caption{\label{fig3} Two-tone spectroscopy (background subtracted) of three devices with $M=20$, $200$ and $1000$. Frequency landscape of the measurement is sketched on the left. A weak tone (green) is used to probe the phase at the resonance of one of the modes that fall within the measurement band of our setup. Another strong drive tone (blue) is swept in frequency. When the drive crosses some resonance of the structure, the Stark shift of the monitored mode due to the self- or cross-Kerr coupling results in a phase response of the probe tone.  Dashed line in the middle panel is a fit to the device model.}
\end{figure*}

Notably, this dependence on $M$ remains extremely weak, and the Kerr nonlinearity of the fundamental mode varies by only about a factor of two in the large range of $M\in[5,200]$. We believe such a weak dependence accounts for the fact that arraying has not brought significant improvement of the dynamic range in parametric amplifiers, despite the significant effort in this direction by the community. For instance, amplifiers made with arrays of 8 SQUIDs \cite{Zhou2014}, 80 SQUIDs \cite{Planat2018} and 20 SNAILs \cite{Frattini2018} have dynamic range comparable or worse than the device consisting of just a single SQUID \cite{Roy2015}. For such comparison to be fair fabrication-wise, we note that the critical currents of the junctions used in all these designs are comparable and lie in the 1--3 $\rm\mu A$ range. Moreover, the dynamic range of Josephson TWPAs \cite{Macklin2015,Planat2019} with $M>2000$ is close to what can be achieved with resonant amplifiers with $M=20$ and careful choice of operating condition \cite{Sivak2019}, and is only one order of magnitude above the best {\it single} SQUID performance \cite{Roy2015} (despite having three orders of magnitude more junctions). Such comparison underlines the issue that arraying does not efficiently dilute the nonlinearity.

\section{Array mode frequencies and Kerr nonlinearities: Experiment\label{sec:Probing the array modes}}

In order to confirm our understanding of the scaling with $M$, we fabricated and tested three devices named A, B and C, with $M\in\{20,200,1000\}$ varying on a logarithmic scale. From Fig.~\ref{fig2}(c),(d) we see that these devices should have mode structure characteristic of regions I, III and IV respectively. 

The devices are fabricated with a single lithography step in a microstrip geometry on a 300-$\mu$m-thick silicon substrate. The ground plane is formed by a 2-$\mu$m-thick silver layer deposited on the back of the wafer. The Josephson junctions are formed by ${\rm Al}/{\rm AlO}_x/\rm{Al}$ layers deposited using the Dolan bridge shadow evaporation process.

The devices are coupled to a $50\,\Omega$ environment via signal and pump ports, as illustrated in Fig.~\ref{fig2}(a). The signal port is strongly coupled and sets the damping rate $\kappa\sim(150-250)\rm\,MHz$ of the resonant modes. In devices A and B this coupling is capacitive, while in device C it is galvanic. The galvanic coupling of device C changes its mode structure compared to the model discussed in Section~\ref{sec:Effect of the array size}, but does not affect the major results of this work, such as the scaling of nonlinearity with $M$. Such a coupling was a necessary design choice to achieve large $\kappa$ in device C. The pump port in each device is  weakly capacitively coupled to a dedicated pump line. The device properties are summarized in Appendix~\ref{App:Description of the device}. To characterize them, we first measure the resonant mode structure.

The resonances can be located with two-tone spectroscopy \cite{Masluk2012,Krupko2018}. The frequency landscape of such a measurement is sketched in Fig.~\ref{fig3} on the left. The first weak tone (green) is used to probe the resonance of one of the modes that fall within the $(4-12)\,\rm GHz$ measurement band of our setup, depicted in Fig.~\ref{figS1}. This tone is applied through the strongly coupled port of the device and measured in reflection. At the same time, the second strong drive tone (blue), applied through the weakly coupled pump port, is swept in frequency (and in power, to compensate for increasing attenuation at higher frequencies). When the drive crosses some resonance of the structure, the Stark shift of a monitored mode due to the self- or cross-Kerr coupling results in a phase response of the probe tone. Moreover, the sign of the phase shift corresponds to the sign of the self- or cross-Kerr nonlinearity that caused this shift. 

The measurement result shown in Fig.~\ref{fig3} demonstrates that device A with $M=20$ is indeed in the ``lumped'' regime, with all higher modes being pushed up in frequency, in agreement with expectations for region I of Fig.~\ref{fig2}(c). Its performance as an amplifier was characterized in detail in Ref.~\cite{Sivak2019}. 

The device B with intermediate $M=200$ displays multiple array modes separated by about $5\,\rm GHz$. The decreased mode spacing at higher frequencies is a hallmark of the approaching ``plasma resonance'' as explained in Section \ref{sec:Effect of the array size}. We fit the frequencies of the array modes of this device to extract the values of the distributed capacitance $C_S=0.11\,\rm pF$ and $C_0=0.70\,\rm fF$ characteristic to our fabrication process, which agree well with the expected design parameters. At higher frequency the deviation from the model of Section \ref{sec:Effect of the array size} (dashed lines) is more apparent, which might arise from the long-range Coulomb interaction between the islands of the array, as explained in Ref.~\cite{Krupko2018}. The second mode of the device B behaves as a regular SNAIL parametric amplifier, and was used in experiment \cite{Hays2019} to perform the single-shot readout of Andreev levels in an InAs nanowire Josephson junction. 

\begin{figure*}
 \includegraphics[width = \figwidthDouble]{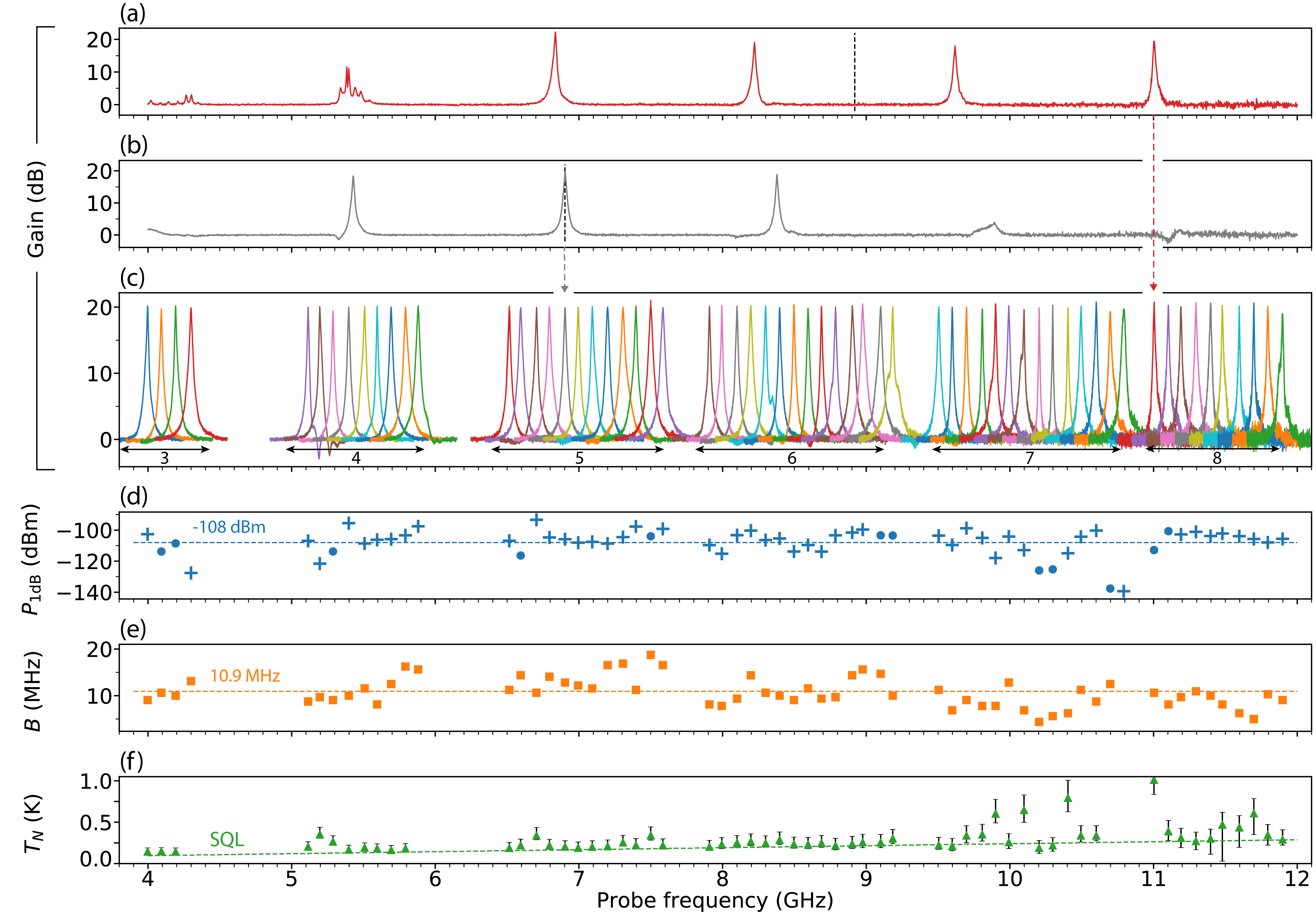}
 \caption{\label{fig4}  (a) A single pump at $\omega_p =2\pi\times 17.84\rm\, GHz$ produces large non-degenerate gain between two signal-idler pairs of modes. (b) A single pump at $\omega_p =2\pi\times 13.81\rm\, GHz$ produces large degenerate and non-degenerate gain. The vertical dashed line is located at $\omega_p/2$. (c) Demonstration of the wide flux tunability of the device. Each gain curve corresponds to a different pumping condition determined by $\Phi$, $\omega_p$, $P_p$. The index of the mode used to produce gain is indicated at the bottom. (d) Saturation power, (e) $3\rm\, dB$ bandwidth and (f) noise temperature, corresponding to the operating points shown in (c). Crosses and circles in (d) correspond to $-1\rm\, dB$ and $+1\rm\, dB$ compression points respectively \cite{footnote}, and we only show the lowest of the two. Horizontal dashed lines in (d) and (e) show the mean value; sloped dashed line in (f) indicates the standard quantum limit $\hbar\omega/2k_B$.}
\end{figure*}

Device C has regularly spaced low-lying modes with $\Delta\omega\approx 1.5 \, \rm GHz$. Due to the similar extent of frequency tunability with flux, the modes of device C are able to almost entirely cover the $(4-12)\,\rm GHz$ frequency range accessible with our measurement setup. The vertical lines of different color visible in the two-tone spectroscopy data likely result from the phase slips in the array when crossing the flux $\Phi=\pm \Phi_0/2$, where $\Phi_0=h/2e$ is the superconducting magnetic flux quantum.

In addition to the resonant mode frequencies of devices A-C, we measure self-Kerr nonlinearities using IMD spectroscopy \cite{Frattini2018,Sivak2019,Weisl2019}. The results are summarized in Table~\ref{tab1} and are shown for reference in Fig.~\ref{fig1}(d); they confirm theoretical expectations. In particular, we are able to reduce the Kerr nonlinearity by an order of magnitude in device C ($M=1000$) over devices A and B ($M=20$ and $200$ respectively). Note that a similar suppression could be achieved in a device with $M=1$ due to lower participation ratio. However, this would come at the expense of flux tunability of only about $100\rm\, MHz$ and inability to pump this device for amplification. 

Having experimentally demonstrated the effect of array size on the mode structure of the device, we now focus on studying the properties of device C belonging to the region IV of Fig.~\ref{fig2}(c). This region looks interesting in application to parametric amplification, although it has not been systematically explored from this perspective before. One example of such a multimode amplifier was characterized in Ref.~\cite{Simoen2015}, where it was demonstrated that two modes of the device can be used for degenerate and non-degenerate gain processes with near quantum-limited added noise. Another example \cite{Winkel2019}, developed independently and in parallel with our work, utilizes pairs of modes (dimers) of  two coupled arrays of SQUIDs to realize a tunable non-degenerate four-wave mixing amplifier.

\section{JAMPA characterization\label{sec:Amplifier characterization}}

In this Section we present detailed characterization of device C as a parametric amplifier. The results of the measurements of various amplifier characteristics are summarized in Fig.~\ref{fig4}. In this experiment, we attempt to obtain $20\rm\, dB$ of gain at regularly spaced operating frequencies separated by approximately $100\rm\, MHz$ within $(4-12)\rm\, GHz$ range. As shown in Fig~\ref{fig4}(c), we were able to cover most of this range except for frequency pockets which are not occupied by any of the modes at any external magnetic flux $\Phi$.  Note that these gaps would disappear when the array size is further increased since the mode spacing decreases as $1/M$.

The measurement is done using an automated search algorithm, which for each desired operating frequency $\omega_{op}$ identifies the flux $\Phi$ at which one of the modes, indexed by $n$, is at resonance near $\omega_{op}$, so that $\omega_n\approx\omega_{op}$. Then, for each combination of modes $(n,m)$, by pumping around $\omega_n+\omega_m$, the algorithm attempts to find the pump power $P_p$ and frequency $\omega_p$ that yields $20\rm\,dB$ of gain at $\omega_{op}$. Thus, some of the gain curves in Fig.~\ref{fig4}(c) correspond to nondegenerate gain ($n\neq m$), as shown in Fig.~\ref{fig4}(a), and some correspond to degenerate gain ($n=m$), as shown in Fig.~\ref{fig4}(b). The dashed line in these panels indicates the location of $\omega_p/2$, which allows to identify if the gain curve is degenerate or nondegenerate. By looking at a wide frequency span it is clear that the same pumping condition that yields $G=20\rm\, dB$ at the desired frequency $\omega_{op}$, also yields large gain in other modes that satisfy the frequency matching condition $\omega_{n'}+\omega_{m'}=\omega_p$, resulting in a comb-like gain profile.

For each operating condition from Fig.~\ref{fig4}(c), we measure the $1\rm\, dB$ compression point $P_{\rm 1 dB}$, $3\rm\, dB$ bandwidth $B$, and noise temperature $T_N$, shown in Fig.~\ref{fig4}(d)-(f). On average, we obtain $B/2\pi=10.9\rm\, MHz$ and $P_{1 \rm \, dB}=-108\rm\, dBm$ with some points as high as $-93\rm\, dBm$. This automated measurement is not optimized and reflects the average amplifier performance, but we expect that by careful fine-tuning of the controls (in particular, pump detuning \cite{Liu2017,Sivak2019}) the above-average performance can be obtained at most operating frequencies. 

The amplifier added noise temperature $T_N$ and $1\rm\, dB$ compression power $P_{1\rm dB}$ were calibrated using a shot noise tunnel junction (SNTJ) \cite{Spietz2003,Spietz2010}, as explained in Appendix~\ref{App:Noise calibration}. For a phase-preserving linear amplifier, the standard quantum limit (SQL) on the added noise, represented with a dashed line in Fig.~\ref{fig4}(f), corresponds to $T_N=\hbar\omega/2k_B$. This half a photon of noise comes from the idler, whose presence is fundamentally necessary to preserve the commutation relations of the amplified operator, as stated by the Caves theorem \cite{Caves1982}. As seen from Fig.~\ref{fig4}(f), the JAMPA closely approaches this fundamental limit on added noise over a wide range of frequencies, which has a practical significance in potential application of JAMPA to qubit readout. More fundamentally, this agreement explicitly demonstrates Caves theorem between 4 and 12 GHz. Note that some points above $9\rm\, GHz$ significantly deviate from the quantum limit, possibly due to the fact that the unselective pump might stimulate spurious conversion processes with unintended modes, leading to existence of multiple idlers and increased added noise. Another possibility is the multistability of the  metapotential created by the parametric pump \cite{Sivak2019, Wustmann2019}. A more sophisticated tuning algorithm might resolve this problem.

Having characterized various aspects of the JAMPA, we move on to discussing its possible applications in circuit QED and ways to improve its performance. 

\section{Discussion and Conclusion\label{sec:Conclusion}}

From a practical point of view, the JAMPA demonstrates equally good performance over a broad tunable bandwidth making it useful in experiments that require sequential, rather than simultaneous measurements of multiple qubits. Fast flux lines or pump switching would be used for such sequential measurements of different channels \cite{Abdo2017}. Using the JAMPA in such a setting can save a large amount of space in a dilution refrigerator compared to using multiple narrow-band parametric amplifiers \cite{Kou2018}. 

In addition, with further reduction of the mode spacing, the comb-like gain profile of JAMPA can become useful for frequency-division multiplexing (FDM) \cite{Jerger2012,Schmitt2014,Heinsoo2018,Kundu2019} of multiple readout signals in a single amplifier. In this technique, to distribute the available bandwidth between multiple communication channels, it is optimal to arrange them at equally spaced frequencies, which makes it suitable for using the JAMPA. 

We also envision more exotic applications of the JAMPA. For example, a squeezing comb can be used to generate continuous-variable cluster states \cite{Grimsmo2017}, which are multimode entangled states that are resources for measurement-based universal quantum computation \cite{VanLoock2007}. Such states can be constructed by coupling various modes of the JAMPA with squeezing and conversion pumps, as was demonstrated for tripartite entanglement generation in Ref.~\cite{SandboChang2018}. 

There is a lot of room for future investigation and improvement. The JAMPA has been developed with the goal of using arraying to suppress the Kerr nonlinearity and thus enhance the amplifier dynamic range. As evident from Fig.~\ref{fig1}(b), the Kerr suppression was effective in this device, although the $P_{\rm 1 dB}$ has not improved significantly over previous SNAIL parametric amplifiers. This is because the saturation power scales quadratically with the coupling rate $\kappa$, which was a few times smaller in the current device than state-of-the-art SPAs. Increasing the coupling can be achieved in several ways, for example by lowering the characteristic impedance $Z_S=390\, \Omega$ of the array. Since ${\Delta\omega}/{\kappa}=({\pi}/{2}){Z_S}/{R}$, lowering the impedance $Z_S$ can lead to a situation in which the comb-like gain profile with multiple resolved peaks turns into a single broad-band gain profile. Understanding this transition is an interesting theoretical problem. Notably, the idea behind a broadband parametric amplifiers from Refs.~\cite{Roy2015,Naaman2017} falls under the same category of using ``blending'' of multiple modes to enhance the dynamic bandwidth. 

This brings us back to the question raised in the Introduction: is there an alternative to the TWPA, which achieves similar performance on important metrics, but has significantly smaller constraints on fabrication? We conjecture that the JAMPA in the ``mode-blending'' regime could potentially serve as such an alternative.

To conclude, we have demonstrated how multiple array modes of a chain of tunable nonlinear mixing elements based on Josephson junctions can be used to dramatically enhance the tunable bandwidth of a Josephson parametric amplifier, while also achieving the state-of-the-art dynamic range and near quantum-limited noise performance.

\section{Acknowledgment\label{sec:acknowledgement }}

We acknowledge the contribution of N.E.~Frattini, A.~Lingenfelter, C.~Ding, W.~Dai and S.O.~Mundhada. JA wishes to acknowledge important discussions with F.~Lecocq. We also acknowledge the Yale Quantum Institute. Any mention of commercial products for information only; it does not imply recommendation or endorsement by NIST. Facilities use is supported by the Yale SEAS clean room and YINQE. This research is supported by AFOSR under Grant No. PFA9550-15-1-0029 and by ARO under Grants No. W911NF- 18-1-0212, No. W911NF-18-1-0020.

\appendix

\section{Description of the devices and measurement setup \label{App:Description of the device}}

The device properties are summarized in Table~\ref{tab1}. The geometric parameters of the SNAIL, fabrication process, chip layout and packaging are similar to that in Appendix A of Ref.~\cite{Sivak2019}. A schematic of the cryogenic microwave measurement setup used in this experiment is shown in Fig.~\ref{figS1}. All measurements were performed using various measurement classes of Keysight PNA-X N5242A network analyzer.

\begin{figure*}
 \includegraphics[width = \figwidthDouble]{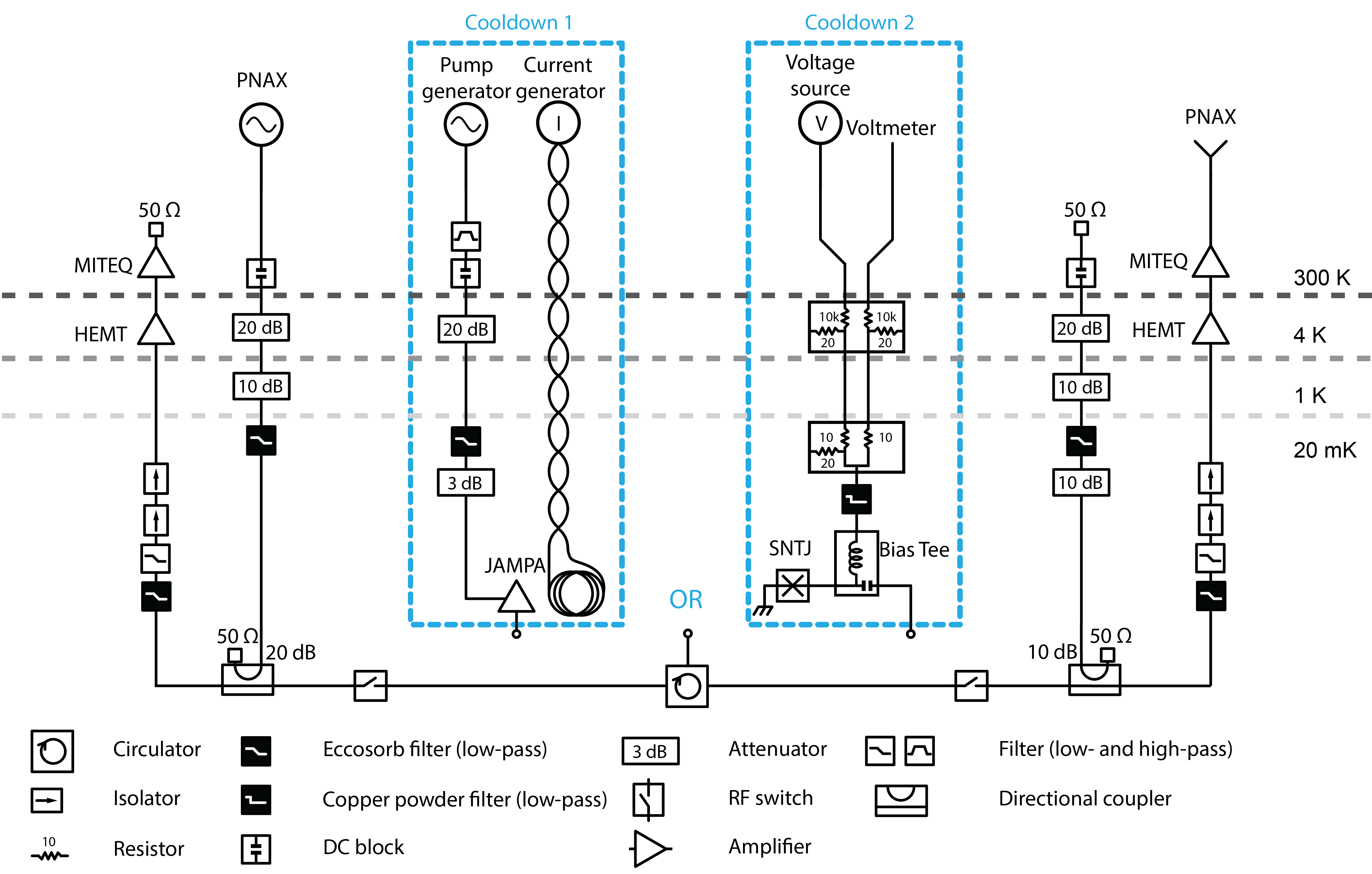}
 \caption{\label{figS1} Schematic of the cryogenic microwave measurement setup used in this experiment. The extra input and output lines and RF switches, are used in other experiments not described in this article. Parts of the setup enclosed in blue boxes were connected to the circulator port in two separate cooldowns.}
\end{figure*}

\newcolumntype{C}{>{\centering\arraybackslash}p{2cm}}
\begin{table*}
\begin{tabular}{|C|C|C|C|C|C|C|C|}
\hline
Device & $M$ & $L_J$ (pH) & Mode, $n$ & $\omega_n/2\pi$  (GHz) & $\kappa_n/2\pi$ (MHz) & $K_{n}/2\pi$ (kHz) \\ \hline \hline
A	& 20 & 38 & 1 & 7.3	& 260 & 25.0\\ \hline
B 	& 200 & 48 & 2 & 10.4 & 150 & 27.5 \\ \hline
\multirow{4}{*}{C} & \multirow{4}{*}{1000} & \multirow{4}{*}{48} & 3 & 4.4 & 150 & 1.0 \\ 
 	&  &  & 4 & 6.1 & 140 & 2.0 \\ 
 	&  &  & 5 & 7.8 & 140 & 3.7 \\ 
 	&  &  & 6 & 9.6 & 140 & 5.0 \\ \hline
\end{tabular}
\caption{\label{tab1} Parameters of three devices studied in this experiment: number of SNAILs ($M$), inductance of the large junctions in the SNAIL ($L_J$), mode number ($n$), resonant frequency ($\omega_n$), coupling rate to the transmission line ($\kappa_n$), and self-Kerr nonlinearity ($K_n$) measured at $\Phi=0$.}
\end{table*}

\section{Eigenmode decomposition and nonlinearities of the JAMPA \label{App:Eigenmode decomposition}}

In this Appendix we derive the resonant mode structure for the system shown in Fig.~\ref{fig2}(a), treating both the transmission line resonator and the array in the continuous approximation. 

\subsection{The SNAIL}

In this section we provide a brief description of the SNAIL properties, and refer to Refs.~\cite{Frattini2018,Sivak2019} for more details. The potential energy of the SNAIL with the junction inductance ratio $\alpha$ can be written as
\begin{align}
U_S(\varphi) = -E_J\left[ \alpha\cos\varphi+3\cos\left(\frac{\varphi_{\rm ext}-\varphi}{3}\right) \right],
\end{align}
where $\varphi$ is the phase across the small junction of the SNAIL, $\varphi_{\rm ext}=\Phi/\phi_0$ and $\phi_0=\hbar/2e$ is the reduced flux quantum. Introducing the generalized flux $\Delta\phi=\phi_0\varphi$ and expanding the potential around the minimum $\phi_{\rm min}$, we obtain 
\begin{align}
U_S = \frac{(\Delta\phi-\phi_{\rm min})^2}{2 L_S(\Phi)}+E_J\sum_{n=3}^{\infty}\frac{c_n(\Phi)}{n!\phi_0^n}(\Delta\phi-\phi_{\rm min})^n,
\end{align}
where $L_S(\Phi)=L_J/c_2(\Phi)$ is the flux-dependent linear inductance and the Taylor expansion coefficients $c_n(\Phi)$ depend only on $\Phi$ and fixed asymmetry coefficient $\alpha$. Note that any other inductive circuit element shown in Fig.~\ref{fig2}(b) can be incorporated into the formalism of this Section by appropriately changing of the expansion coefficients $c_n$ and the location of the potential minimum $\phi_{\rm min}$.

In the continuous model, we replace $\Delta\phi\to a\partial_x\phi$ and absorb the gradiometric flux offset into the definition of $\phi$: $\phi-\frac{x}{a}\phi_{\rm min}\to\phi$.

\subsection{Resonant frequencies \label{App: Resonant frequencies}}

Let us denote the capacitance and inductance per unit length of the transmission line resonator as $c_{r}$ and $\ell_{r}$ respectively. Then the characteristic impedance is given by $Z_{c}=\sqrt{\ell_{r}/c_{r}}$, and the phase velocity is given by $v_{r}=1/\sqrt{\ell_{r}c_{r}}$. Similarly, for the SNAIL array with unit cell of size $a$, the capacitance to ground and inductance per unit length are $c_{0}=C_{0}/a$ and $\ell_{S}(\Phi)=L_{S}(\Phi)/a$ respectively, and the shunting capacitance of the unit cell is $c_{S}=C_{S}a$. With these definitions, we can write the Lagrangian of a linearized system as
\begin{align}
{\cal L} & =\bigg(\int_{-\frac{d_{1}}{2}}^{-\frac{d_{0}}{2}}+\int_{\frac{d_{0}}{2}}^{\frac{d_{1}}{2}}\bigg)\bigg[\frac{c_{r}}{2}(\partial_{t}\phi)^{2}-\frac{1}{2\ell_{r}}(\partial_{x}\phi)^{2}\bigg]dx\nonumber \\
+ & \int_{-\frac{d_{0}}{2}}^{\frac{d_{0}}{2}}\bigg[\frac{c_{0}}{2}(\partial_{t}\phi)^{2}+\frac{c_{S}}{2}(\partial_{x}\partial_{t}\phi)^{2}-\frac{1}{2\ell_{S}}(\partial_{x}\phi)^{2}\bigg]dx,\label{eq:continuous Lagrangian}
\end{align}
where $d_{0}=Ma$ is the total array size and $d_{1}=Ma+2d_{r}$ is the total size of the device. By varying the action for this Lagrangian, we can find the Lagrange equations of motion for the flux field $\phi(x,t)$:\\
$\circ$ In the transmission line resonator
\begin{equation}
-c_{r}\partial_{t}^{2}\phi+\frac{1}{\ell_{r}}\partial_{x}^{2}\phi=0.\label{eq:wave eq in microstrip}
\end{equation}
$\circ$ In the SNAIL array
\begin{equation}
-c_{0}\partial_{t}^{2}\phi+c_{S}\partial_{x}^{2}\partial_{t}^{2}\phi+\frac{1}{\ell_{S}}\partial_{x}^{2}\phi=0.\label{eq:wave eq in array}
\end{equation}

In addition, this procedure yields the continuity conditions at the various boundaries of the system: (i) zero-current boundary condition at the ends of the transmission line
\begin{align}
\partial_{x}\phi\big|_{x=-\frac{d_{1}}{2}} & =0\nonumber \\
\partial_{x}\phi\big|_{x=+\frac{d_{1}}{2}} & =0\label{eq:boundary1}
\end{align}
(ii) Current continuity condition at the points connecting the transmission line resonator and the array
\begin{align}
\frac{1}{\ell_{r}}\partial_{x}\phi\bigg|_{x=-\frac{d_{0}}{2}-0} & =\bigg(\frac{1}{\ell_{S}}\partial_{x}\phi+c_{S}\partial_{x}\partial_{t}^{2}\phi\bigg)\bigg|_{x=-\frac{d_{0}}{2}+0},\nonumber \\
\frac{1}{\ell_{r}}\partial_{x}\phi\bigg|_{x=+\frac{d_{0}}{2}+0} & =\bigg(\frac{1}{\ell_{S}}\partial_{x}\phi+c_{S}\partial_{x}\partial_{t}^{2}\phi\bigg)\bigg|_{x=+\frac{d_{0}}{2}-0}.
\end{align}
(iii) Flux (or voltage) continuity condition at the points connecting the transmission line resonator and the array
\begin{align}
\phi\big|_{x=-\frac{d_{0}}{2}-0} & =\phi\big|_{x=-\frac{d_{0}}{2}+0},\nonumber \\
\phi\big|_{x=+\frac{d_{0}}{2}+0} & =\phi\big|_{x=+\frac{d_{0}}{2}-0}.\label{eq:boundary6}
\end{align}

Because we have neglected the nonlinearity of the SNAIL, all the obtained equations of motion are linear and can be solved analytically. We will look for a harmonic solution corresponding to the eigenfrequency $\omega$ in the piece-wise form
\begin{equation}
\phi=\begin{cases}
\begin{array}{cc}
A_{-}\cos k_{r}x+B_{-}\sin k_{r}x, & x<-d_{0}/2\\
A_{S}\cos k_{S}x+B_{S}\sin k_{S}x, & |x|<d_{0}/2\\
A_{+}\cos k_{r}x+B_{+}\sin k_{r}x, & x>d_{0}/2
\end{array}\end{cases}\label{eq: ansatz solution}
\end{equation}
where the wave-vectors $k_{r}$ (in the resonator) and $k_{S}$ (in the array) are related to the frequency $\omega$ via
\begin{align}
\omega^{2} & =\frac{1}{\ell_{r}c_{r}}k_{r}^{2},\\
\omega^{2} & =\frac{1}{\ell_{S}c_{S}}\frac{k_{S}^{2}}{k_{S}^{2}+\frac{c_{0}}{c_{S}}},
\end{align}
which follows from Eqs.~\eqref{eq:wave eq in microstrip} and \eqref{eq:wave eq in array} in Fourier domain. By substituting the ansatz solution given by Eq.~\eqref{eq: ansatz solution} into Eqs.~\eqref{eq:boundary1}-\eqref{eq:boundary6}, we obtain a system of six linear equations to determine the amplitudes
$\left(\begin{array}{cccccc}
A_{-} & B_{-} & A_{S} & B_{S} & A_{+} & B_{+}\end{array}\right)^{T}$.  The eigenmode frequencies of the system can be determined by requiring that the determinant of this system of equations is zero, while the eigenvectors provide the spatial profile of the eigenmode flux distribution. The equation for the determinant of the system can be factorized and solved separately for the odd and even modes
\begin{align}
\tan\bigg(\frac{d_{r}\omega}{v_{r}}\bigg)\tan\bigg(\frac{M\omega}{2\omega_{0}\sqrt{1-\frac{\omega^{2}}{\omega_{p}^{2}}}}\bigg) & =\frac{Z_{c}}{Z_{S}}\sqrt{1-\frac{\omega^{2}}{\omega_{p}^{2}}},\label{eq:odd sols}\\
\tan\bigg(\frac{d_{r}\omega}{v_{r}}\bigg)\cot\bigg(\frac{M\omega}{2\omega_{0}\sqrt{1-\frac{\omega^{2}}{\omega_{p}^{2}}}}\bigg) & =-\frac{Z_{c}}{Z_{S}}\sqrt{1-\frac{\omega^{2}}{\omega_{p}^{2}}},\label{eq:even sols}
\end{align}
where we have introduced the ``plasma frequency'' $\omega_{p}=1/\sqrt{L_{S}C_{S}}$, the characteristic impedance of the SNAIL transmission line $Z_{S}=\sqrt{L_{S}/C_{S}}$, and $\omega_{0}=1/\sqrt{L_{S}C_{0}}$. Equations \eqref{eq:odd sols} and \eqref{eq:even sols} can be solved numerically to yield the resonance frequencies $\omega_{n}$ of the array modes. In order to produce Fig.~\ref{fig2}(c), we fix the fundamental mode at the desired operating frequency $\omega_{op}$, and the corresponding length of the resonator pads has to be determined from
\begin{equation}
d_{r}=\frac{v_{r}}{\omega_{op}}\arctan\bigg\{\frac{Z_{c}}{Z_{S}}\sqrt{1-\frac{\omega_{op}^{2}}{\omega_{p}^{2}}}\cot\bigg(\frac{M\omega_{op}}{2\omega_{0}\sqrt{1-\frac{\omega_{op}^{2}}{\omega_{p}^{2}}}}\bigg)\bigg\}.
\end{equation}

The critical $M=M_{c}$ after which the resonator pads shrink to zero and we lose $d_{r}$ as a frequency-controlling knob, is given by
\begin{equation}
M_{c}=\pi\frac{\omega_{0}}{\omega_{op}}\sqrt{1-\frac{\omega_{op}^{2}}{\omega_{p}^{2}}}.
\end{equation}

For $M>M_{c}$, we can simplify the expressions \eqref{eq:odd sols}-\eqref{eq:even sols}, and obtain the closed-form solution for the array mode frequencies
\begin{equation}
\omega_{n}=\frac{\omega_{p}}{\sqrt{1+\left(\frac{M\omega_{p}}{\pi n\omega_{0}}\right)^{2}}}.\label{eq:dispersion_relation_JAMPA}
\end{equation}

It is clear from this expression that the frequencies $\omega_n$ of the low-lying modes are approximately equally spaced with $\Delta\omega\approx\pi\omega_0/M$.

\subsection{Participation ratio}

It is helpful to know how the participation of the array changes with $M$, as it controls characteristic features in the resonant frequency and self-Kerr dependence on $M$. We can use the solution $\phi(x)$ of the linearized system to define the inductive energy participation ratio (EPR) as

\begin{align}
p = \frac{\int_{-\frac{d_{0}}{2}}^{\frac{d_{0}}{2}}\frac{(\partial_{x}\phi)^{2}}{2\ell_{S}}dx}{
\left(\int_{-\frac{d_{1}}{2}}^{-\frac{d_{0}}{2}}+\int_{\frac{d_{0}}{2}}^{\frac{d_{1}}{2}}\right)\frac{(\partial_{x}\phi)^{2}}{2\ell_{r}}dx+
\int_{-\frac{d_{0}}{2}}^{\frac{d_{0}}{2}}\frac{(\partial_{x}\phi)^{2}}{2\ell_{S}}dx}.
\label{eq:EPR}
\end{align}

\begin{figure*}
 \includegraphics[width = \figwidthDouble]{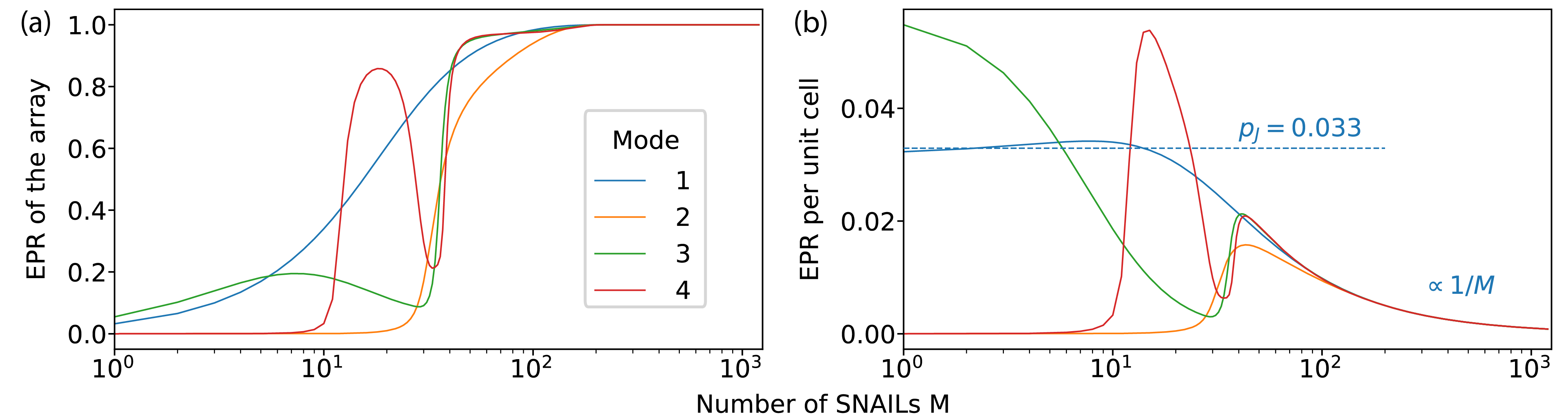}
 \caption{\label{figS2} (a) Inductive energy participation ratio $p$ of the array in the total inductive energy of the mode, based on the quadratic approximation. (b) EPR scaled by the number of unit cells in the array.}
\end{figure*}

The dependence of the EPR on the number of unit cells in the array $M$ is shown in Fig.~\ref{figS2}. In general, this dependence is non-monotonous, showing that the mode structure changes in a nontrivial way.

We can identify a few features that are nevertheless easy to correlate with Fig.~\ref{fig2}(c),(d). In particular, the linear increase of the self-Kerr nonlinearity at small $M$, as prediced by Eq.~\eqref{propto M}, relies on the EPR of the unit cell being independent of $M$. This condition is confirmed in Fig.~\ref{figS2}(b), where the saturation of $p/M$ to $p_J\approx0.033$ is apparent. In contrast, at large $M$ the participation ratio becomes 1 by definition, and its per-unit value decreases as $1/M$, leading to dilution of the  nonlinearity in the JAMPA. Note that in reality $p$ would saturate to a value smaller than $1$ because of the spurious inductance of the leads within the circuit of a SNAIL unit cell. Our model does not account for this effect, although it becomes significant when the geometric inductance of the SNAIL loop is comparable to $L_J$.

The separation of the diagram in Fig.~\ref{fig2}(c) into four regions roughly corresponds to where EPR per unit cell (I) stays approximately constant, (II) reduces to half its asymptotic value, (III) crossovers into a $1/M$ dependence, and (IV) becomes strictly equal to $1/M$.

\subsection{Self-Kerr nonlinearities}

After solving the linearized system, we proceed to study the nonlinearities. In general, the Hamiltonian obtained after eigenmode decomposition contains nonlinear couplings of all orders and between all modes. Therefore it can be written as
\begin{align}
H & =\sum_{n}\omega_{n}a_{n}^{\dagger}a_{n}+\sum_{ijk}g_{3}^{(ijk)}(a_{i}+a_{i}^{\dagger})(a_{j}+a_{j}^{\dagger})(a_{k}+a_{k}^{\dagger})\nonumber \\
+ & \sum_{ijkl}g_{4}^{(ijkl)}(a_{i}+a_{i}^{\dagger})(a_{j}+a_{j}^{\dagger})(a_{k}+a_{k}^{\dagger})(a_{l}+a_{l}^{\dagger})+...
\end{align}

Our goal is to establish the characteristic scaling of nonlinearities with $M$, and therefore we will make a few simplifying assumptions. First, we focus on $\Phi=0$. In this case all odd nonlinearities vanish due to the inversion symmetry, and the amplifier cannot be used in three-wave mixing mode at this particular external magnetic flux. However, the method presented here can be extended easily to calculate any coupling $g_{3}^{(ijk)}$ at arbitrary flux. Second, we will only calculate the self-Kerr nonlinearities $K_{n}\equiv12g_{4}^{(nnnn)}$. We expect that this linear participation ratio based method would need to be significantly modified in order to calculate the Kerr nonlinearity at non-zero flux, due to corrections from the third-order coefficient in the SNAIL potential energy, as was shown in Refs.~ \cite{Frattini2018,Sivak2019}. However, we note that this method can still be applied easily at non-zero flux for systems in which the Kerr nonlinearity is the lowest order nonlinearity, such as arrays of Josephson junctions or SQUIDs. 

According to the black-box quantization (BBQ) method \cite{Nigg2012}, after the linearized problem is solved, the nonlinearity can be simply applied to the linear flux distribution found in Eq.~\eqref{eq: ansatz solution}. For the SNAIL array at $\Phi=0$, the fourth-order nonlinear contribution to the Hamiltonian is given by 
\begin{equation}
U_{{\rm NL}}=\frac{1}{a}\int_{-d_{0}/2}^{+d_{0}/2}\frac{c_{4}}{4!L_{J}\phi_{0}^{2}}(a\partial_{x}\phi)^{4}dx.
\end{equation}

Next, we plug in the flux distribution \eqref{eq: ansatz solution}  and integrate over $x$, which leads to quartic nonlinear couplings between the amplitudes $A_{S}$ and $B_{S}$ of various modes. Since we are only interested in the self-Kerr nonlinearities $K_n$, we select only the terms in the potential energy $U_{\rm NL}$ that are related to the mode $n$. These potential energy terms have the form
\begin{widetext}
\begin{align}
U_{n}=\frac{c_{4}}{4!L_{J}\phi_{0}^{2}}\frac{(ak_{S})^{3}}{16}\left[6(A_{S}^{2}+B_{S}^{2})^{2}(d_{0}k_{S})+8(-A_{S}^{4}+B_{S}^{4})\sin(d_{0}k_{S})+(A_{S}^{4}-6A_{S}^{2}B_{S}^{2}+B_{S}^{4})\sin(2d_{0}k_{S})\right].
\end{align}
\end{widetext}

Note that the amplitudes $A_{S}$ and $B_{S}$ are not independent, and the relation between them is established when the eigenvalue problem is solved for the system of equations (\ref{eq:boundary1})-(\ref{eq:boundary6}). 

In general this is a cumbersome albeit straightforward procedure which is much easier done numerically, but we present here the outline of actions for the easily solvable case of $d_{r}=0$. In this case the dispersion relation has the form (\ref{eq:dispersion_relation_JAMPA}). The open boundary condition \eqref{eq:boundary1} leads to the dimensional quantization of the wave-vector $k_{S}^{(n)}=\frac{\pi}{a}\frac{n}{M}$, and the eigenmodes have a simple structure with $A_{S}=0$ for odd modes and $B_{S}=0$ for even modes. Given this, the contribution to the Hamiltonian from mode $n$ simplifies to
\begin{align}
H_{n} & =\bigg(\frac{C_{0}M}{4}+\frac{C_{S}M(ak_{S}^{(n)})^{2}}{4}\bigg)\dot{D}^{2}+\frac{(ak_{S}^{(n)})^{2}M}{4L_{S}}D^{2}\nonumber \\
 & +\frac{c_{4}}{4!L_{S}\phi_{0}^{2}}\frac{3}{8}(ak_{S}^{(n)})^{4}MD^{4}
\end{align}
where $D$ can be either $A_{S}$ or $B_{S}$ depending on the parity of the mode. After applying canonical quantization, this can be transformed to
\begin{equation}
H_{n}=\omega_{n}a_{n}^{\dagger}a_{n}+\frac{1}{12}K_n(a_{n}+a_{n}^{\dagger})^{4},\label{eq: H_n}
\end{equation}
with $\omega_{n}$ given by (\ref{eq:dispersion_relation_JAMPA}) and $K_n$ given by 
\begin{equation}
K_n=\frac{3}{16}\frac{c_{4}}{c_{2}}\frac{L_{S}}{M\phi_{0}^{2}}\omega_{n}^{2}.
\end{equation}

Note that the Hamiltonian is not simply the sum of the terms of type \eqref{eq: H_n} since it also contains cross-Kerr couplings between all modes.

Although the more general case with $d_r\neq0$ can also be treated analytically, the obtained expressions are cumbersome and do not provide much insight. Instead, we solve the general case numerically and present the results in Fig.~\ref{fig2}(d). For numerics, we use the following realistic system parameters: $\omega_{op}=8\rm\, GHz$, $C_0=0.71\rm\, fF$, $C_S=0.11\rm\, pF$, $L_S=110\rm\, pH$, $Z_c=46\rm\, \Omega$, $v_r=1.2\times 10^8\rm\, m/s$.

\section{Noise calibration \label{App:Noise calibration}}

We calibrated the noise temperature of the JAMPA $T_N$ and its $1\rm\, dB$ compression point $P_{\rm 1dB}$, shown in Fig.~\ref{fig4}(d),(f), using a shot noise tunnel junction (SNTJ) \cite{Spietz2003,Spietz2010}. The noise emitted from the SNTJ, and amplified by the full cryogenic and room temperature measurement chain, has the following power spectrum:
\widetext
\begin{align}
P_N(\omega) = G_{\rm sys}(\omega)k_B B\left[ T_{sys}(\omega) + \frac{1}{2}\left( \frac{eV+\hbar\omega}{2k_B}\right)\coth\left( \frac{eV+\hbar\omega}{2k_BT}\right) + \frac{1}{2}\left( \frac{eV-\hbar\omega}{2k_B}\right)\coth\left( \frac{eV-\hbar\omega}{2k_BT}\right)  \right],
\end{align}
\twocolumngrid
\noindent where $G_{sys}(\omega)$ is the frequency-dependent system gain, $T_{sys}(\omega)$ is the frequency-dependent system noise temperature (without the parametric amplifier), $T$ is the junction temperature, $V$ is the voltage applied across the junction, $B$ is the resolution bandwidth over which noise is collected,  and $k_B$ and $\hbar$ are Boltzmann and Planck constants. We use the high-voltage limit of this expression, 
\begin{align}
P_N(\omega) = G_{\rm sys}(\omega)k_B B\left[ T_{sys}(\omega) + \frac{e|V|}{2k_B}\  \right],\label{eq:noise power}
\end{align}
which allows for simple extraction of $G_{\rm sys}(\omega)$ and $T_{sys}(\omega)$ using linear fits. 

To extract $T_N$, we mount the JAMPA in place of the SNTJ and measure the noise visibility ratio (NVR). $T_N$ is extracted from
\begin{align}
{\rm NVR} = \frac{G_{sys}(T_{sys}+G(T_Q+T_{N}))}{G_{sys}(T_{sys}+T_Q)},\label{nvr}
\end{align}
where $G$ is the gain of JAMPA and $T_Q=\hbar\omega/2k_B$. The standard quantum limit (SQL) corresponds to $T_{N}=T_Q$. 

\begin{figure}
 \includegraphics[width = \figwidth]{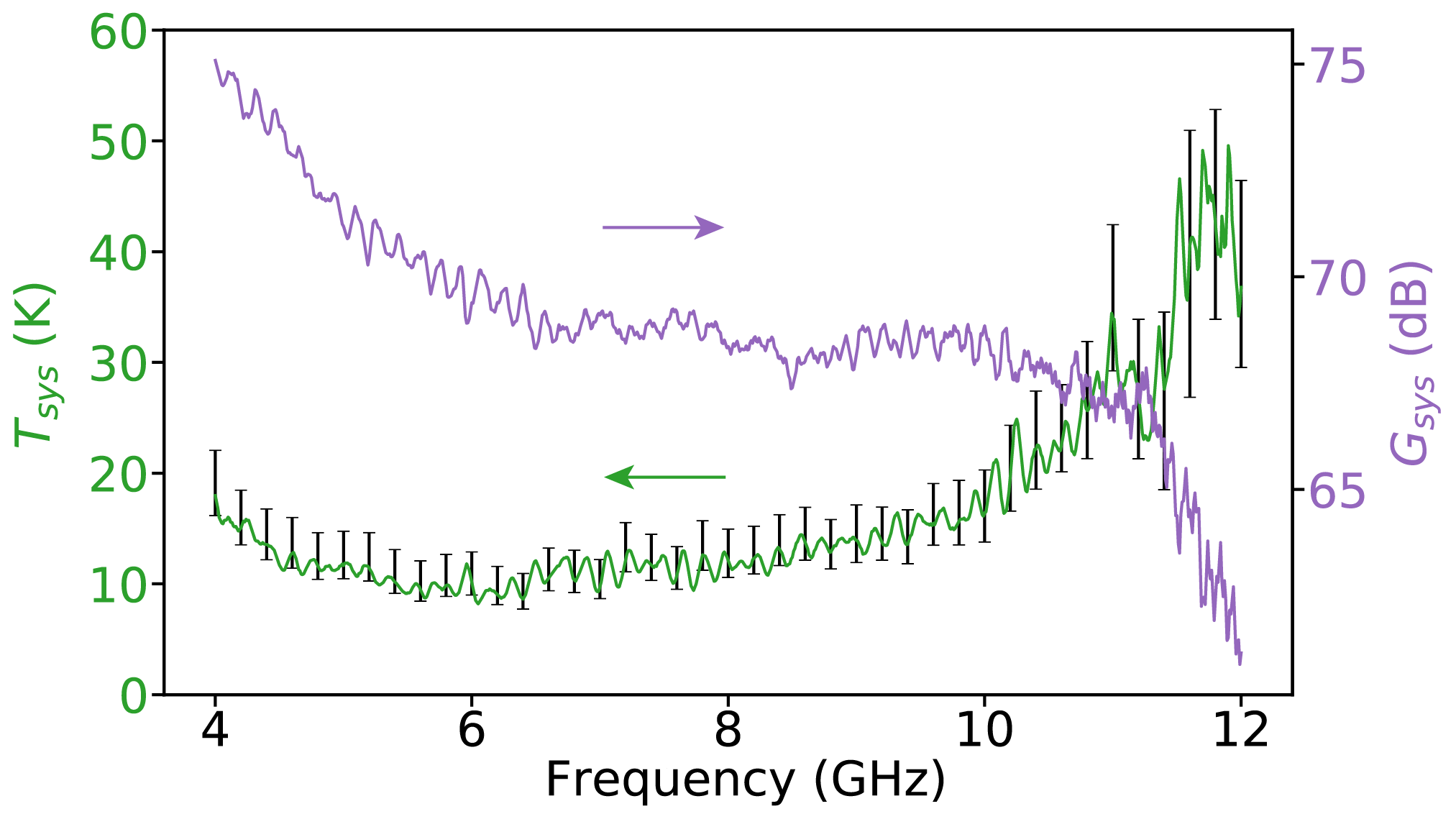}
 \caption{\label{figS3} System noise temperature $T_{sys}(\omega)$ and gain $G_{sys}(\omega)$ of the amplification chain measured using SNTJ mounted in place of JAMPA.}
\end{figure}

Following this outline, we measured $T_{sys}(\omega)$ and $G_{\rm sys}(\omega)$, shown in Fig.~\ref{figS3}, in a separate cooldown, after mounting the SNTJ in the same place as JAMPA. The total gain of the chain in this case includes contributions from the high-electron-mobility transistor (HEMT) amplifier mounted at the 4K stage of the dilution refrigerator and the MITEQ amplifier at room temperature. The increased system noise compared to the factory specifications of the HEMT (model LNF-LNC4-16) results from the attenuation between the HEMT and the SNTJ, and is typical of such setups \cite{Roy2015,Simoen2015,SandboChang2018}.

The error bars in $T_{sys}$ and $T_N$ result from three major possible systematic errors: 
\begin{enumerate}[label=(\roman*)]
\item The finite $S_{11}$ of the SNTJ means that not all noise power is collected due to the impedance mismatch. This effect leads to an overestimation of $T_N$ which gets worse at higher frequency when approaching the on-chip resonance of the SNTJ at about $11.5\rm\, GHz$.
\item The possible miscalibration of the voltage across the SNTJ due to the change in the ratio of the resistive voltage divider at low temperature. This error is symmetric and estimated to be about $\pm10\%$.
\item The shift of the calibration reference plane due to the bias tee used with the SNTJ. This effect was studied in detail in \cite{Su-WeiChang2016}: it can be modeled as a beamsplitter that mixes the voltage dependent noise of the SNTJ with (presumably quantum) noise of the bias tee. This causes SNTJ to report a higher system noise temperature and smaller gain according to
\begin{equation}
G_{sys}' = \eta G_{sys}, \quad  T_{sys}' = \frac{1}{\eta}T_{sys}+\frac{1-\eta}{\eta}T_Q
\end{equation}
where $\eta\approx -0.6\rm\, dB$ is the insertion loss of the bias tee measured at room temperature. We account for the finite $\eta$ in the noise calibration. We note however that $\eta$ can increase when the bias tee is cooled down which would increase $T_N$ by up to $10\%$ (corresponding to $\eta=1$).
\end{enumerate}
Other sources of error, such as the statistical error of the fit and the error due to measuring the SNTJ in a separate cooldown, are insignificant in comparison.

Knowing $G_{sys}(\omega)$, we can also calibrate the absolute power at the device plane, which determines $P_{1\rm dB}$. We find that this value is on average $2.3\rm\, dB$ higher than that extracted with the knowledge of the input line attenuation when the dilution refrigerator is warm; this difference is consistent with the fact that the line attenuation reduces slightly at low temperature.

\bibliographystyle{apsrev_longbib}
\bibliography{library}

\end{document}